\pdfoutput=1
\documentclass[11pt]{article}
\usepackage[T1]{fontenc}
\usepackage[utf8]{inputenc}
\usepackage{lmodern}
\usepackage{microtype}
\usepackage[margin=1.1in]{geometry}
\usepackage{graphicx}
\usepackage{booktabs}
\usepackage[numbers,sort&compress]{natbib}
\usepackage[colorlinks=true,linkcolor=blue!45!black,citecolor=blue!45!black,urlcolor=blue!45!black]{hyperref}
\usepackage{xcolor}
\newcommand{\partsubtitle}[1]{\begin{center}\itshape #1\end{center}}
\newcommand{\uarr}{\ensuremath{\rightarrow}}
\newcommand{\uin}{\ensuremath{\in}}
\newcommand{\uDelta}{\ensuremath{\Delta}}
\newcommand{\uleq}{\ensuremath{\le}}
\newcommand{\ugeq}{\ensuremath{\ge}}
\newcommand{\utau}{\ensuremath{\tau}}
\newcommand{\utausq}{\ensuremath{\tau^{2}}}

\title{Where Reliability Lives: Experimental Localisation of Behavioural Properties in an Agent System}
\author{Timothy Marsden, Matthew Collecutt and James Marsden\\ Taniwha AI}
\date{August 2026}
\begin{document}
\maketitle
\begin{abstract}

Every reliability claim about an agentic system implicitly locates a
property somewhere: in the model, or in the machinery around it. If a
system is said to be incapable of an invalid action, it matters greatly
whether that incapacity lives in a prompt, in learned model behaviour, or
in a boundary the model cannot reach — and in practice this is usually
read off an architecture diagram rather than tested. We built a system in
which the location of such properties can be experimentally probed, and
intervened separately on cognition and on institutional epistemic
mechanisms.

The subject is a persistent simulated settlement: autonomous inhabitants
act, an authoritative append-only ledger adjudicates every attempted act
against world state, and accepted history is the only reality. Mind,
institution and world were separated before any experiment was designed.

Holding cognition fixed, we intervened on the institution's epistemic
mechanisms — evidence provenance, belief availability, physical-evidence
legibility. Repairing provenance alone moved false attribution from 44 to
4 of 107 marked-face verdicts. These experiments were preregistered before
their instruments existed, and refuted our central registered prediction
twice, in opposite directions. A preregistered falsifier then supplied the
one input the decomposition's geometry had structurally denied the belief
channel — a staged veridical first-hand witness — and the channel's marginal
value, non-positive throughout the witness-free phases, turned positive:
nine of eleven seeds improved, at zero added false attribution.

Holding institutional enforcement fixed, we intervened on cognition four
ways: ablating the native minds' machinery, killing and resetting minds
mid-task, replacing the entire native cognition with a frozen
frontier-LLM panel, and corrupting beliefs with trusted false testimony.
Behaviour changed dramatically each time — throughput collapsed and
recovered, and one identical falsehood cost each trusting run roughly nine
hundred futile actions while costing the distrusting arm none. Five
pre-declared properties did not move in any tested trajectory: accepted
reality stayed singular, invalid attempts were refused with typed reasons,
duties survived the processes that held them, no work was accepted twice,
and no false completion was ever accepted — 2,581 completion claims from
the substituted panel, none false.

In the cell preregistered as most dangerous to this design, the
substituted panel outperformed our own architecture on refusal
handling; the invariants survived it.

Our claim is limited to this setting: in this one institution, measured
behavioural properties were separable from substantial changes to
cognition, and that separation was established by direct intervention. This is one designed world, under
interventions that degrade or replace cognition; it does not test a
capable agent searching for a route around the institution, and it
establishes nothing about a population of institutions. What it does show
is that where a system-level behavioural property resides can be treated
as an experimental question rather than an architectural assertion.

\end{abstract}
\section{Introduction}

Discussions of reliability in agentic AI systems tend to treat it as
one quantity with one home: a better model behaves better, and less
or worse information should produce more uncertainty. Both intuitions
are natural defaults — confidence read as a proxy for epistemic
health, abstention as a signal that a system knows what it does not
know, and the model as the seat of all of it. An older engineering
tradition locates reliability elsewhere: in transaction managers,
type systems, ledgers and institutions that make certain bad states
unrepresentable regardless of the actor's quality. Modern agentic
systems contain both layers — a cognitive model, and a growing
apparatus of policy, memory, provenance, tool constraints and
adjudication above which the model acts. Which layer is responsible
for which reliability property is seldom experimentally separated in
deployed systems, whose layers are rarely built so that one can be
intervened on while the other holds still.

This paper reports a programme that performs that separation from
both directions, in one controlled setting. The instrument is an
executable institution whose mind/institution/world separation, typed
seams and enforcement locations were architectural commitments made
before any of these interventions was designed — the experiments
intervene on a pre-declared boundary rather than a partition fitted
to results.
Part I holds cognition fixed and intervenes on the institution's
epistemic mechanisms; reliability there changes in
mechanism-specific, causally attributable ways, including in ways
that refuted our own registered predictions. Part II holds
institutional enforcement fixed and intervenes radically on cognition
— ablating it, killing it mid-task, replacing it wholesale with a
frozen language-model panel, and corrupting it through trusted
testimony — and asks which pre-specified properties survive.

The five properties, fixed in advance and scored against the world's
own record throughout Part II:

1. \textbf{Singular accepted reality} — one authoritative accepted
   history, never forked or privately redefined;
2. \textbf{Typed refusal of invalid acts} — every inadmissible attempt is
   refused with a machine-readable reason, not silently dropped;
3. \textbf{Duty recovery from institutional history} — obligations are
   reconstructible from the institution's record alone, without
   surviving private state;
4. \textbf{Zero duplicate accepted work} — the same consumable work is
   never accepted twice;
5. \textbf{Zero false completions} — a claim of completed duty that the
   world's own progress record contradicts is never accepted.

Stated as one claim: \textbf{in this executable institution, institutional
and cognitive reliability properties were experimentally separable}
— reliability is not monolithic; some of its properties moved with
cognition, some with the institution, and whether each tested
property survived cognition-layer intervention was determined
experimentally. Two guards govern every use of that
sentence in this paper. First, the experiments do not show that
cognition is unimportant: cognition mattered enormously for
throughput, efficiency, belief hygiene, refusal handling and cost,
and Part II reports those differences in full, including
where they are adverse to our own architecture. Second, the two
programmes intervene on opposite sides of the same pre-declared
boundary; they are \emph{not}
a crossed cognition \texttimes{} institution factorial, and we do not present
them as one. Whether the separation survives when both sides move at
once is precisely the factorial reserved as future work (§8).

The absence of a learned model from Part I is the control, not a gap
in relevance: Part I's interventions act on exactly the institutional
and epistemic machinery that modern agentic systems layer above the
cognitive model, while the cognitive architecture itself is held
fixed. Part II then varies
the cognition — including replacing it with a frontier language
model — against the same enforcement. The two parts address different
limitations of the design. Part I alone could be read as an
architecture
measuring its own reflexes; Part II shows the institutional
guarantees holding under every cognition change we made, including
full substitution. Conversely, Part II alone could
be read as designed invariants surviving by construction; Part I
shows that intervening on institutional mechanisms moves real
outcomes, causally and reproducibly.

Our contributions:

1. \textbf{A causal result} (§4.1): a single-commit ablation of one
   provenance rule changes collective attribution quality by a large,
   replicated margin.
2. \textbf{An observability result} (§4.2): provenance corruption can
   cross a failure threshold while remaining invisible on the answer
   plane — legible only at the mechanism carrying the evidence, while
   a monitored health indicator moves in the reassuring direction.
3. \textbf{A channel-separation result, with its registered boundary test}
   (§4.3–4.5): the institutional belief channel's marginal value,
   measured against a no-belief counterfactual at every level of
   physical-evidence quality, is non-positive throughout the
   witness-free decomposition regime; whether evidence-starved cases
   resolve to honest "unexplained" or to false accusation is causally
   shifted by the channel's availability. The preregistered falsifier
   (§4.5) then handed the channel its strongest input — a staged
   veridical first-hand witness — and its marginal value turned
   positive (9 of
   11 seeds, zero false names in 32 fired verdicts): the channel's
   measured value moved with what it carried, not with the channel's
   presence as such.
4. \textbf{An ablation result} (§5.1): removing the native mind's
   perception-admission gate collapses the mind — and the damage
   propagates into public outcomes — while the five properties hold.
5. \textbf{A substitution result} (§5.3): replacing the entire native
   cognition with a frozen context-only frontier-LLM panel preserves
   all five properties on the same seeds; in the preregistered
   dangerous cell the panel meets contention and stands down where
   the native cognition retries roughly forty-one times — an adverse
   finding for the native architecture.
6. \textbf{A corruption result} (§5.4): a declared trust relation causally
   determines whether identical false testimony enters operative
   belief; believing minds pay roughly nine hundred futile acts per
   run-week while the institution refuses all 5,455 resulting
   attempts and accepts none.
7. \textbf{The registered predictions that failed} (§7): our central
   registered prediction was refuted twice in opposite directions; a
   mid-programme invariant failed its own registered test; a frozen
   prediction about speculative discipline failed; and a comparator
   beat our architecture in the cell preregistered as most dangerous
   to the design.

We do not claim these results generalise beyond the regimes we tested
(§8). This paper measures intervention effects within one frozen
institution; it estimates nothing about a population of institutions.
We claim the results are real within it, that they were obtained
under preregistered designs and scoring procedures intended to limit
post-hoc interpretation, and that
they motivate measuring reliability properties at the mechanisms that
enforce them rather than inferring them solely from the behaviour or
confidence of the cognitive model.

\begin{figure}[t]
\centering
\includegraphics[width=\linewidth]{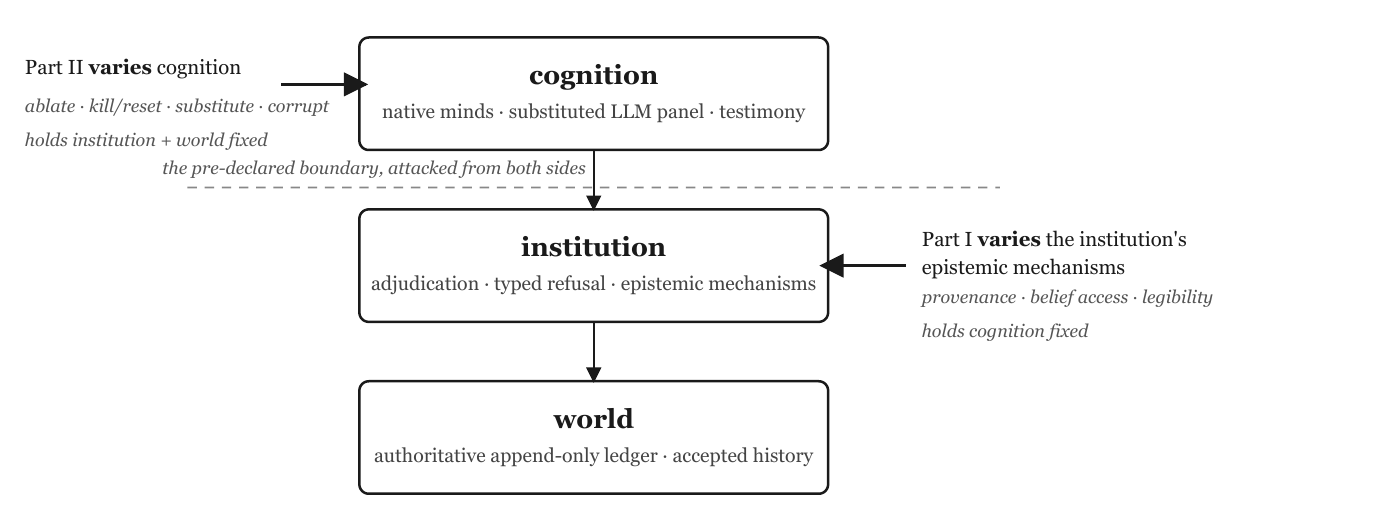}
\caption{the pre-declared boundary. Part II varies the cognition layer and holds the institution and world fixed; Part I holds cognition fixed and varies the institution's epistemic mechanisms. The five properties of §1 are enforced below the cognition layer.}
\label{fig:1}
\end{figure}

\section{The instrument}

\textbf{The world and its institution.} Copperhollow is a persistent
simulated settlement whose physical history — extraction, structural
wear, water, collapses, injuries — is computed lawfully from state
(no event dice; failures are threshold crossings under load) and
recorded in an authoritative, append-only institutional ledger. The
ledger is the experiment's ground truth: every verdict and every
acceptance is scored against conserved world truth the inhabitants
cannot alter, so "correct," "false," and "unresolved" are properties
of a record, not of a rater. The institution — the ledger institution
throughout — adjudicates every attempted act: an admissible act is
\emph{accepted} into the single history; an inadmissible one is \emph{refused}
on the wire with a typed, machine-readable reason. Acceptance is
adjudicated against world state, not against the actor's beliefs or
identity. Duties, once disclosed, are part of institutional history;
leases and reservations guard contended resources; completed work is
recorded with correlations identifying the consumed subject 1:1,
which is what makes duplicate-acceptance and false-completion scoring
exact rather than heuristic.

\textbf{The native minds.} Inhabitants are deterministic agents with
individual, typed, provenance-carrying belief stores. Perception is
deterministic and skill-gated. Beliefs carry provenance metadata;
testimony moves beliefs between minds through an explicit channel. We
use \emph{belief} throughout in the standard agent-systems sense — an
internally stored proposition available for action selection and
adjudication; no claim about consciousness or human-like belief is
intended. The native architecture includes a perception-admission
gate deciding which observations become durable beliefs;
evidence-maintained retirement; predictive frames grading
expectations against events; a curiosity pathway; and
sleep/consolidation. Two known absences are load-bearing for Part II:
action licensing consults belief content but filters neither source
nor weight, and no decision path consumes refusal evidence.

\textbf{Two senses of "provenance," separated up front.} Part I intervenes
on the \emph{institution's} evidence-provenance rule — a stamp on
institutional evidence plus the adjudication step that consumes it.
Part II's native minds carry \emph{private} provenance-carrying belief
state — cognition-side machinery a substituted model lacks. The same
word, two referents, on opposite sides of the boundary; every use
below is one or the other, and the separability claim depends on not
conflating them.

\textbf{Two experimental regimes, one world.} Part I runs a frozen
sabotage-tribunal regime: a seeded camp in which saboteurs cut
mine-face supports at night, faces collapse days later, and a fixed
adjudication ladder (the "tribunal") produces per-verdict outcomes
scored against the ledger's true cause — \textbf{correct} (sabotage
identified), \textbf{false attribution} (an innocent blamed), or
\textbf{unexplained} (measured, never discarded). The ladder consults, in
order of directness: witnessed sabotage; physical tool marks gated by
examiner skill and by \emph{surprise} (a face already believed failing is
not examined closely); witnessed pillar-robbery; believed-marginal
maintenance; a long-unserviced alarm. One geometric fact calibrates
every Part I result about beliefs: the saboteurs strike unwitnessed,
at night — the strongest possible belief-channel signal, direct
eyewitness identification, is structurally absent throughout the
decomposition phases, and the witnessed-sabotage rung fired in \textbf{zero
of all 253 decomposition runs, anchors included}. Every claim about
the belief channel's value in §4.1–4.4 is scoped to this witness-free
geometry (§8); §4.5 reports the preregistered boundary test that
supplies exactly the missing signal. The provenance rule at Part I's
centre: sign beliefs carry the band of the reading they replaced, and
the tribunal's surprise gate consumes it — an alarm that \emph{jumped}
overnight re-opens close examination.

Part II runs the settlement's duty economy: five minds holding
standing duties (drawing water, winning ore, hauling, depositing)
against the same institution. Its substituted cognition (the panel)
is a context-only driver, one loop per mind, each calling a single
frozen frontier LLM route (\texttt{deepseek-v4-flash}, temperature 0,
JSON-object responses) with the institution's disclosures and events
as context — no persistence, so killing the driver destroys the
panel's entire "memory" by construction. Every claim this paper makes
about the panel is scoped to this one comparator at its one measured
cadence coupling (§5.3, §8). An external "oracle" can
deliver testimony into minds through an ordinary inlet carrying
source provenance; a mind's declared trust relation to it is a
scalar: positive trust admits at that weight through the normal
admission path, a bypass arm commits directly, negative trust rejects
delivery and journals the rejection.

\begin{figure}[t]
\centering
\includegraphics[width=\linewidth]{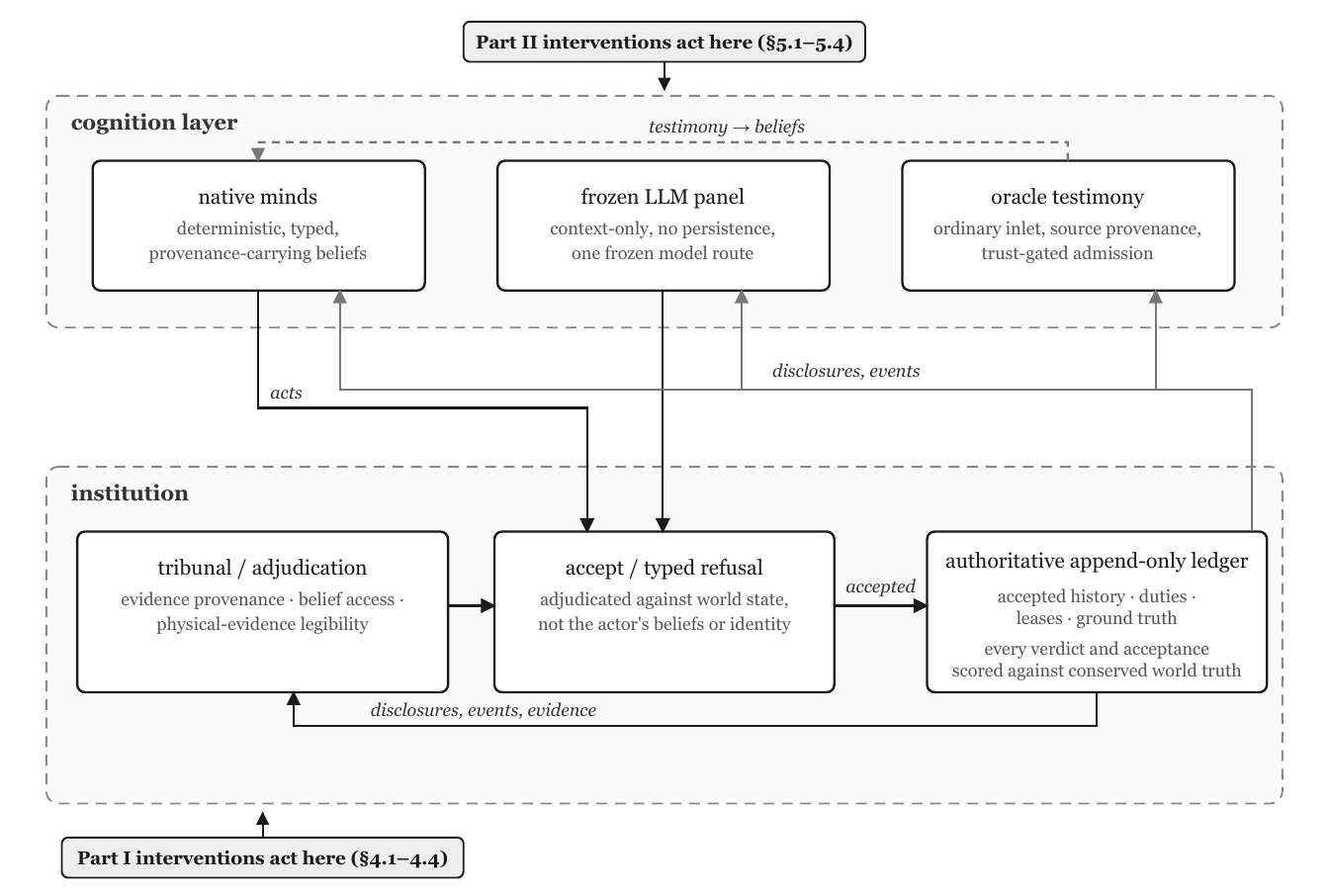}
\caption{the instrument, and where each intervention acts. Part I's treatments act on the tribunal's epistemic mechanisms; Part II's act on the cognition layer; every attempted act is adjudicated against world state into a single accepted history.}
\label{fig:2}
\end{figure}

\section{Method: replication and preregistration against a frozen boundary}

The programme used two related experimental forms, and the order of
operations in each is load-bearing. The first (§4.1) is a controlled
replication and causal isolation of an already-observed repair: its
arms are existing commits, and the anchor result was known before the
contrast was designed — what was fixed in advance there is the
replication protocol, the seed set, and the identity gates, not the
direction of the effect. The remaining experiments reported here —
the degradation grids of §4.2–4.4, the witnessed-strike boundary test
of §4.5, and all of Part II — were
preregistered before their intervention instruments existed. The
shared protocol:

1. \textbf{Designs pinned before the instruments existed} (all experiments
   except §4.1's replication). Intervention surfaces, levels, seeds,
   primary outcomes, registered rival outcomes, exclusion (VOID)
   rules and analyses were committed to version-controlled design
   documents before the switches implementing the interventions were
   written; interpretation freezes were written before any number
   existed.
2. \textbf{Declared intervention surfaces.} The degradation treatments of
   §4.2–4.4 are single
   guarded code sites controlled by environment variables, off by
   default, with deterministic per-entity draws; where a draw's key
   was shared across seeds in the first degradation experiment, the
   artefact was
   identified, disclosed, and corrected to seed-salted draws in all
   later degradation experiments (§8). Part II treatments are declared arms
   (ablations, a seeded kill, an accepted-model set matched exactly,
   a frozen scripted insult delivered at a fixed game-minute).
3. \textbf{Disposability and identity gates.} The degradation switches at
   null
   settings were required to reproduce shipped behaviour exactly at a
   pinned anchor seed, and measured endpoints to reproduce prior
   measured arms exactly. E3 and E5 carried a frozen subject identity at three levels — source commit, compiled wheel digest, runtime image digest — enforced at launch by a gate that refused on mismatch, because a source-level pin alone was shown unable to distinguish two subjects differing only in compiled behaviour. E1 and E2 recorded no image digest and enforced no such gate; their execution images have since been identified from surviving execution artefacts, which is provenance recovered after the fact and not a control exercised at collection. Admission was fail-closed behind identity, completeness and
   world-invariance gates, and an uninterpretable trajectory is VOID
   — an apparatus outcome, never a subject result.
4. \textbf{Frozen scoring.} All scoring gates and registries were frozen
   before collection; no threshold in this paper was chosen after
   seeing data.
5. \textbf{Disclosed defects.} The programme's own failures are part of
   its record: a contaminated first collection destroyed before
   inspection; three cells voided and recollected under identical
   seeds; five trajectories voided on a model-provenance ambiguity
   and replaced under a tightened exactly-matched gate; retired runs
   retained under labels, never deleted.
6. \textbf{Sealed artefacts, independent recomputation.} Every run grid
   and score file is hash-addressed; every headline number in this
   paper was independently recomputed from sealed artefacts before
   use — for this paper, every score file was regenerated
   byte-identically from raw trajectories under the frozen scorers,
   and six prose figures in internal reports that failed
   recomputation are corrected here (§7.1).

\textbf{Statistical treatment.} Part I's unit of replication is the seed
trajectory: one pinned anchor seed plus ten fresh seeds, identical
across its experiments; pooled verdict counts are descriptive, and
every effect claim is also stated at seed level as a paired contrast
(direction counts, medians, ranges). Part II's cells are n=3
trajectories under frozen seeds and conditions, treated as controlled
trajectories, not samples from a population; direction-consistency is
reported descriptively. We deliberately report no significance tests,
for a different reason in each part. In Part I, runs are
deterministic given seed, and the seed sets are finite, declared in
advance, and exhaustively evaluated — inferential statistics over
hypothetical seed populations would imply a sampling model we do not
possess. In Part II the cells are small fixed grids of controlled
trajectories rather than samples, and the panel arm is not
seed-deterministic at all — a live model route couples to world
timing — so its claims are stated descriptively per trajectory, never
inferentially. Where a claim is exact ("zero duplicates,"
"zero false completions," "all refusals typed"), it is a count over
the stated denominator, scored against the world's own record.

\section{Part I — hold cognition fixed, intervene on institutional epistemic mechanisms}
\partsubtitle{Epistemic channels of an executable institution}

Part I's minds are the deterministic natives throughout; no learned
model participates in any Part I decision loop. Whatever changed
here, no learned model changed — because none was present.

Part I comprises the witness-free decomposition (§4.1–4.4; 253 runs,
anchors included) and its preregistered boundary test (§4.5; 88 fresh
runs, 341 in all) — designed, frozen and collected after the
decomposition's results were sealed, at the exact missing case the
decomposition's own scope guard named.

\subsection{One provenance rule, ablated (causality)}

\textbf{Design.} Two arms, eleven seeds each, differing by exactly one
commit: the arm \emph{without} the provenance rule of §2 (sign beliefs do
not remember what they replaced; the surprise gate treats every alarm
as standing) versus the arm \emph{with} it. A third arm at the current
development head verified that subsequent, unrelated commits leave
the result untouched. Both anchor-seed controls reproduced their
previously pinned in-vivo results exactly.

\textbf{Result.} On 107 verdicts concerning genuinely sabotaged faces per
arm:

\begin{center}
\small
\begin{tabular}{lccc}
\toprule
\textbf{arm} & \textbf{correct} & \textbf{unexplained} & \textbf{false attribution} \\
\midrule
without the rule & 35 (32.7\%) & 28 (26.2\%) & 44 (41.1\%) \\
with the rule & 75 (70.1\%) & 28 (26.2\%) & 4 (3.7\%) \\
current head & 75 (70.1\%) & 28 (26.2\%) & 4 (3.7\%) \\
\bottomrule
\end{tabular}
\end{center}

The repair converts 40 of 44 false accusations into correct
attributions. (Denominators, once: the pooled 44 comprises 39
fresh-seed false attributions plus 5 at the anchor seed; the repaired
40 = 35 fresh + 5 anchor. Seed-level statements in this paper concern
the ten fresh seeds.) The unexplained count is \emph{exactly invariant}:
the rule did not buy correctness through caution — it added
discriminating evidence, and wrong verdicts became right ones. The
current head is verdict-identical to the repair commit at every seed.

\emph{Seed level (paired):} false attribution never increased in any seed
(10/10 \uleq{} 0). Baseline miscarriages occur in six of ten seeds; the
repair eliminates them entirely in five (median paired change \textminus{}3
across the ten paired seeds, range \textminus{}8 to 0, with correct attribution
rising correspondingly), and leaves one seed's four false
attributions untouched in both arms — that seed's miscarriages are
insensitive to this treatment (they are repaired by a different
intervention; §4.3). One seed produced no marked-face verdicts. The
pooled contrast is thus carried by the five provenance-sensitive
seeds, and no seed moved adversely.

\subsection{Graded provenance corruption (failure presentation)}

\textbf{Design.} The provenance stamp carried with probability
p \uin{} {1.0, 0.75, 0.5, 0.25, 0.0}; 55 runs; deterministic draws (the
one experiment with the shared-key draw design; §8); consumer left in
place; both endpoints anchored to §4.1's measured arms.

\textbf{Result} (ten fresh seeds; 102 marked-face verdicts per level):

\begin{center}
\small
\begin{tabular}{lccc}
\toprule
\textbf{p} & \textbf{correct} & \textbf{unexplained} & \textbf{false attribution} \\
\midrule
1.00 & 70 (68.6\%) & 28 (27.5\%) & 4 (3.9\%) \\
0.75 & 70 (68.6\%) & 28 (27.5\%) & 4 (3.9\%) \\
0.50 & 35 (34.3\%) & 28 (27.5\%) & 39 (38.2\%) \\
0.25 & 35 (34.3\%) & 28 (27.5\%) & 39 (38.2\%) \\
0.00 & 35 (34.3\%) & 28 (27.5\%) & 39 (38.2\%) \\
\bottomrule
\end{tabular}
\end{center}

The response is a step, not a curve: full effect above a threshold,
full ablation at and below it, with p = 0 reproducing §4.1's ablated
arm exactly. Two observations matter more than the threshold's
location (an artefact of the shared-key draw; §8). First, the
unresolved rate never moves. Second — the observability result,
stated at its full measured width, on both planes. On the \textbf{answer
plane}, nothing diagnosed the failure: verdict volume, decisiveness,
and the behavioural-invariants gate were unchanged at every level
while false attribution increased roughly tenfold. On the \textbf{mechanism
plane}, the failure was loud: the surprise gate's own trace — the
count of sudden-alarm classifications — tracked the treatment exactly
(fresh-seed pooled 35 / 35 / 0 / 0 / 0 across the five levels), and
the one monitored indicator that did move, a verdict-grounds
sufficiency check, moved in the \emph{reassuring} direction — becoming
decisive in three provenance-sensitive seeds as the corruption took
hold. The institution became epistemically worse while an ordinary
health indicator looked better; the defect was legible at the
mechanism carrying the evidence, and at none of the monitored
answer-plane indicators. In this experiment, epistemic health was
measurable at the enforcing mechanism and not through the monitored
answer-plane indicators — whether the failure was observable at all
depended on which plane was instrumented. \emph{Seed level (paired):} the
high-versus-low-legibility contrast in false attribution is never
adverse in any seed (10/10 \ugeq{} 0) and strictly positive in exactly
§4.1's five provenance-sensitive seeds (median +3 across the ten
paired seeds, range 0 to +8); the sixth miscarriage-bearing seed's
false attributions persist unchanged at every level of this
treatment.

\subsection{Degrading institutional belief access (channel separation)}

\textbf{Design.} Each belief row consulted by the tribunal's evidence walk
is visible with probability p \uin{} {1.0, 0.75, 0.5, 0.25, 0.0};
seed-salted draws; the world, perception, and belief \emph{formation}
untouched — only the adjudication's access is degraded. 55 runs.
Registered rival outcomes included a non-monotonicity via the
surprise gate (hiding an examiner's own standing alarm can \emph{open}
close examination).

\textbf{Result} (ten fresh seeds; 102 verdicts per level):

\begin{center}
\small
\begin{tabular}{lccc}
\toprule
\textbf{p} & \textbf{correct} & \textbf{unexplained} & \textbf{false attribution} \\
\midrule
1.00 & 70 (68.6\%) & 28 (27.5\%) & 4 (3.9\%) \\
0.75 & 70 (68.6\%) & 29 (28.4\%) & 3 (2.9\%) \\
0.50 & 73 (71.6\%) & 28 (27.5\%) & 1 (1.0\%) \\
0.25 & 75 (73.5\%) & 27 (26.5\%) & 0 (0.0\%) \\
0.00 & 75 (73.5\%) & 27 (26.5\%) & 0 (0.0\%) \\
\bottomrule
\end{tabular}
\end{center}

The blinded tribunal is, in pooled terms, \emph{better}: correct
attribution rises and false attribution falls monotonically as
belief-evidence disappears, while abstention stays within its narrow
full-evidence band (26.5–28.4\%) at every level and is lowest at full
removal (27 vs 28). The registered non-monotonicity materialised as a
monotone improvement: hidden standing alarms re-open examination, and
hidden believed-maintenance records starve the false-blame ladder of
innocent candidates — while complete physical evidence continues to
carry correct verdicts.

\emph{Seed level (paired, full removal vs full availability):} the
seed-level statement is \textbf{never harmful, occasionally helpful} — no seed
worsened on either measure (10/10), and the strict improvement is
concentrated (strictly better in 1/10 seeds; median paired change 0,
ranges 0 to +5 correct, \textminus{}4 to 0 false). The concentration
corresponds to the mechanism identified in the artefact, which
supports the set-identity
exactly: the one seed that strictly improves under removal is
precisely the one seed with residual false attribution at full
availability — the same seed whose four miscarriages §4.1's
provenance repair could not touch (false 4 \uarr{} 0 here). Jointly, the
two interventions account for all 44 baseline false attributions: 40
are provenance-sensitive (§4.1) and the remaining 4 are eliminated by
removing belief access. The channel's one unique upside in this
ladder — an eyewitness naming the actual saboteur — never fired in
any of the decomposition's 253 runs, anchors included: a structural
property of the staged night-strike scenario (§2, §8), which makes
the realised accounting one-sided: systematic costs, no realised
benefit. The practical significance is not that belief channels are
harmful; it is that \textbf{channel value is empirically measurable and can
differ substantially from intuitive expectation}. In this regime the
measured contribution of institutional belief access was
non-positive; the registered witnessed-strike test (§4.5) later
supplied the missing case, and did reverse it.

\subsection{Physical evidence \texttimes{} belief availability (resolution)}

\textbf{Design.} The discriminating question — \emph{does the belief channel
supply unique signal when physical evidence fails?} — requires the
no-belief counterfactual at every level of physical evidence. A 5 \texttimes{} 2
factorial: examiner-side legibility of physical tool marks
\uin{} {1.0, 0.75, 0.5, 0.25, 0.0} crossed with the §4.3 switch at its
endpoints; the ledger's raw state — and therefore scoring — untouched
by construction; 110 runs; seed-salted draws; the primary registered
object the interaction.

\textbf{Result} (ten fresh seeds; 102 verdicts per cell):

\begin{center}
\small
\begin{tabular}{lcc}
\toprule
\textbf{marks} & \textbf{beliefs ON — correct/unexpl./false} & \textbf{beliefs OFF — correct/unexpl./false} \\
\midrule
1.00 & 70 / 28 / 4 & 75 / 27 / 0 \\
0.75 & 55 / 32 / 15 & 59 / 43 / 0 \\
0.50 & 35 / 37 / 30 & 39 / 63 / 0 \\
0.25 & 14 / 39 / 49 & 18 / 84 / 0 \\
0.00 & 0 / 42 / 60 & 0 / \textbf{102} / 0 \\
\bottomrule
\end{tabular}
\end{center}

\begin{figure}[t]
\centering
\includegraphics[width=0.8\linewidth]{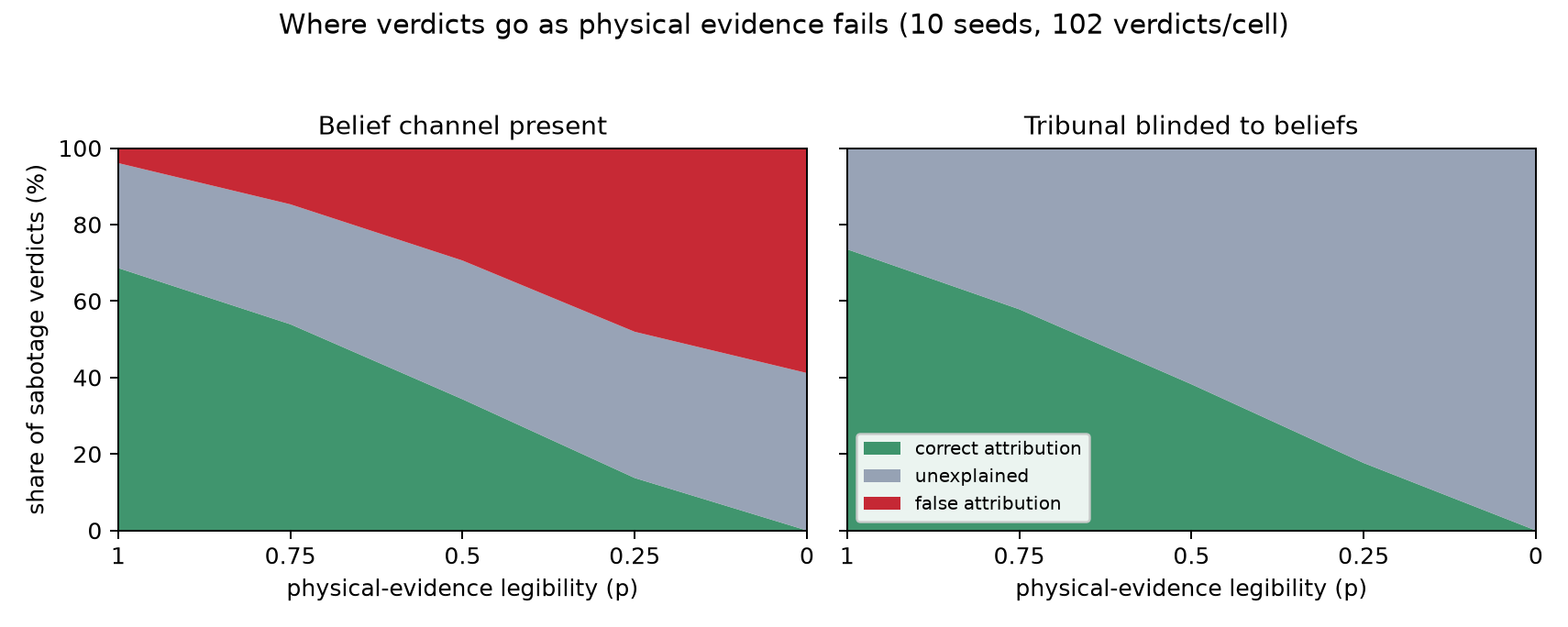}
\caption{as physical-evidence legibility falls, the belief-present tribunal converts evidential darkness into false attribution (red), while the blinded tribunal converts it into honest unresolved verdicts (grey), with zero false attribution anywhere in that arm.}
\label{fig:3}
\end{figure}

Three findings, walked through Figure 3's two panels. \textbf{(a)} The
stable \textasciitilde{}27\% unresolved rate of §4.1–4.3 was not institutional
geometry: it was the evidence-sufficiency floor under full physical
evidence. Blinded to beliefs, the institution's abstention tracks
physical evidence monotonically — to 100\% at zero legibility — with
\emph{zero} false attributions across all 510 verdicts of that arm.
\textbf{(b)} With beliefs present, part of the identical evidential
darkness is converted into false blame instead (4 \uarr{} 60): abstention
rises in both arms as evidence fails, but only the blinded
institution expresses \emph{all} of its evidential loss as honest
unresolved verdicts — with the belief channel present, the channel
fills part of the silence with false attribution. \textbf{(c)} The belief
channel's marginal value in this witness-free factorial is
non-positive at every level of physical evidence
(\uDelta{}correct = \textminus{}5, \textminus{}4, \textminus{}4, \textminus{}4, 0) and its marginal harm grows
monotonically as physical evidence fails (\uDelta{}false = +4, +15, +30, +49,
+60).

\emph{Seed level (paired, beliefs-on \textminus{} beliefs-off):} the interaction is
the programme's most robust effect. \uDelta{}false \ugeq{} 0 in 10/10 seeds at
every legibility level, strictly positive in 8/10 seeds at low
legibility, with the per-seed median rising 0 \uarr{} 2 \uarr{} 3.5 \uarr{} 5.5 \uarr{} 7
across levels — a per-seed dose–response. The blinded arm's
unexplained count is non-decreasing across levels in 10/10 seeds; the
sighted arm's false attribution likewise in 10/10.

Summarising the Part I decomposition: whether
the institution reaches a verdict responds causally to
physical-evidence availability (varied here); within the fixed
examiner population, skill gates access to that evidence (identified
mechanistically, not varied); and under degraded physical evidence in
this witness-free regime, whether lost evidentiary support is
expressed as unresolved uncertainty or converted into false
attribution depends strongly — and causally, under this intervention
— on whether the institutional belief channel remains available.
§4.5 supplies the witness this regime structurally lacks.

\subsection{The registered boundary test: a veridical first-hand witness (the falsifier)}

\textbf{Design.} §4.3–4.4's scope guard named its own missing case: the
belief channel never received the one input for which its ladder rung
exists. The witnessed-strike variant supplies it, and was
preregistered, adversarially reviewed, and frozen — design and built
instrument pinned together by content digest — before any treated run
existed. The intervention is a \textbf{post-decision veridical witness
fold}: with the crime held fixed, exactly one additional eligible
miner (a deterministic, seed-salted draw from the rostered crew,
excluding anyone already naturally present) receives the shipped
first-hand witness fold for the cut, applied after the saboteur's
decision has returned and reading no physical value, so it can change
neither whether nor where the strike lands nor any physical fact the
ledger records about the cut. What it does change is the cut record's
declared witness metadata — exactly one added witness fold — and that
single declared difference is what the causal-boundary gates below
adjudicate per twin. The fold is veridical by construction — it carries
the true actor and face the ledger records, never a fabrication. The
staging is declared: this stages the \emph{existence} of a witness, not an
embodied observer the world's own dynamics produced; whether the
world affords such witnesses is the named successor experiment's
question (§8). No learned model participates.

Grid: witnessed \uin{} {0, 1} \texttimes{} marks legibility \uin{} {1.0, 0.0} \texttimes{} tribunal
\uin{} {sighted, blinded} \texttimes{} the decomposition's 11 seeds = 88 fresh runs,
twinned by seed; the 44 untreated controls were freshly executed
against the frozen instrument, not reused from §4.4, so every
contrast is between twins that differ in exactly one declared thing.
Denominators: 107 marked-face verdicts per cell (11 seeds, pinned
anchor included).

\textbf{The causal boundary, adjudicated per twin.} Four gates,
preregistered: decision identity through the treated strike; physical
identity of the treated cut with only the declared witness metadata
normalised out; exactly one additional declared witness fold relative
to the twin; and the standing gate-error screen. \textbf{All 44 twins pass
all gates; zero runs were voided.} Beyond what the gates require,
the entire saboteur decision stream was identical in every twin: the
treatment changed adjudication without changing the treated crime or
any saboteur decision in the 44 collected twins. Both untreated
anchor cells reproduced the frozen pre-collection gate receipts
byte-identically — a frozen verdict array re-derived by live
re-execution.

\textbf{Result.} The primary contrast was frozen before collection: the
paired difference-in-differences in correctly \emph{named} attribution at
zero marks legibility — the count of rung-one verdicts naming the
ledger's actual cutter, witnessed minus unwitnessed, sighted minus
blinded, within seed. The registered support rule required a positive
mean and a positive difference in at least 6 of 11 seeds; the
collection returned \textbf{mean +1.455 with 9 of 11 seeds positive}, and
the hypothesis is supported at exactly that width. The seeds are the
enumerated decomposition set, not a sample; no significance test is
applied, and the per-seed vector (+2, +1, +1, +1, +1, +2, +3, +2, 0,
+3, 0) is the evidence — seeds in ascending non-anchor order, the
pinned anchor seed last.

Secondaries: the sighted-arm effect is
\textbf{+16 correctly named attributions at each marks level} with the
identical per-seed vector at both — the witness's value did not
depend on physical-marks legibility in this grid; \textbf{zero false names
in all 32 fired verdicts} across both levels; every deciding row
read first-hand by the observational provenance field added for this
study (zero hearsay — independently confirming §2's account that no
rumour path can write the deciding row at this subject); under zero
legibility the witness also displaced false attribution (65 \uarr{} 55 of
107), converting part of the evidential darkness §4.4 showed the
channel filling with false blame into correct attribution instead;
the blinded column is zero throughout, as the frozen switch
definition predicts. In 2 of 11 seeds a witness was seated and the
rung still never fired — the registered reach-failure rival requires
the injected witness to later be a structural casualty at that same
face, and in those seeds the conjunction did not occur: the rival's
mechanism, visible at seed granularity, without carrying the outcome.
In 8 of 176 treated strikes no eligible witness existed; those
strikes stayed unwitnessed and were recorded, never retried.

\textbf{What this establishes, at width.} In the decomposition the channel
carried priors and absences, and its measured value was non-positive;
supplied a first-hand veridical witness, the same channel's value
turned positive with no measured false naming. Across the tested
regimes the channel's adjudicative value changed with what it
carried. Evidentiary content and provenance condition that value
jointly; their independent contributions are not isolated by this
design, and no claim is made that provenance alone controls the sign.

\section{Part II — hold institutional enforcement fixed, intervene on cognition}
\partsubtitle{Durable institutions, replaceable minds}

Part II asks the inverse question: with the institution frozen, what
do the five properties of §1 owe to the cognition above them?
"Replaceable" in this part means exactly that property list, never
overall capability.

\subsection{Ablate the mind's machinery}

Twenty-one trajectories: the full native architecture and six
ablations, three runs each, seven simulated days per run. The ruled
unit of claim for the headline ablation is the compound
perception-admission seam — the ablation removes novelty control,
bounded intake and the admission path together, so the defensible
claim is the ruling's own: \emph{"allowing every observable fact to become
durable cognitive state makes this persistent agent substantially
less capable of acting effectively"} — with no sub-mechanism
attribution inside the entangled collapse.

Removing the admission gate collapsed the mind. Belief stores
exploded from 817 \textpm{} 242 to 44,021 \textpm{} 20,760 (53.9-fold;
perception-domain records 308 \uarr{} 43,879); accepted acts fell
16.7 \textpm{} 1.2 to 5.7 \textpm{} 1.2 with non-overlapping ranges ([16,18] vs
[5,7]); frame-building collapsed 96.2\% (233,171 \uarr{} 8,757); consolidation
reached a mean of one of five minds per run; violation rates tripled
(non-overlapping ranges); and the flooded minds formed \textbf{zero}
beliefs about their own live accepted work, where controls formed
12–18 per run — every one, in the ruling's exact wording, \emph{"100\% of
graded live work-history beliefs were supported by recorder-visible
accepted events."} The damage propagated into public outcomes
through the week's water-adequacy turn — a public institutional
adjudication whose outcome is computed from the accepted water draws
on the ledger, nothing else: all
three ablated worlds resolved it
\texttt{defaulted, low 5} against the controls' \texttt{defaulted, low 3} —
within-condition invariant, between-condition different, so a
mind-side ablation reached authoritative public history through fewer
accepted draws.

The other ablations bound the architecture's middle: timer-based
forgetting traded stability (accepted acts ranged [7,16]); never
forgetting held action at this horizon while accumulating 68.5\% more
current-state perception records (519 vs 308; total stores
436 \textpm{} 9 < 817 \textpm{} 242 < 1,149 \textpm{} 17); removing predictive frames left a
quiet week's action intact while removing the graded error signal
those frames exist to produce; removing sleep was null at this
horizon. Cross-cutting: in all 21 trajectories there were zero
refusal events — in these quiet weeks, well-fitted duties produced no
inadmissible attempt at all — and the five properties of §1 were
never touched. The mind was made substantially less capable; nothing
it did or failed to do forked accepted reality, and its founded-era
inheritance was invariant under every ablation (exactly 56
founded-era claims excluded in every one of the 21 runs).

\subsection{Perturb reality; kill and reset the minds}

Forty-eight trajectories, sixteen cells, under a frozen contrast map:
engineered contradiction and disappearance, mid-task process death,
conflicting testimony, speculative material under sleep, resource
contention with degraded refusal feedback. The ratified framing is
binding and we adopt it verbatim: the headline is not "six mechanisms
passed" — \emph{"the constitutional architecture survived every stress;
several cognitive mechanisms worked; one failed badly; one failed to
distinguish; and one produced an unexpected causal effect not yet
mechanistically explained."} The ruled statement of what the
experiment supports:

\begin{quote}
a separately authoritative institution can support persistent
agents whose private epistemic states diverge, disappear,
reconstruct, speculate incorrectly and attempt inadmissible acts —
without requiring private consensus and without allowing those
private states to redefine accepted reality.
\end{quote}

\textbf{Restart (property 3 under stress).} Minds killed at a fixed
instant mid-undertaking (utsc 65,604), reattached ten game-minutes
later as fresh sessions. In all three runs: duties re-disclosed from
institutional history, founded histories re-inherited, first
post-restart accepted act at 65,624 — twenty game-minutes after death
— and \textbf{zero duplicate accepted work across the boundary}, scored on
the consumed-consumable basis. The ruled sentence: \emph{"continuity of
agency did not require continuity of the private process"} — the past
came from the institution, not from ghost state.

\textbf{Divergence (property 1 under stress).} Under conflicting seeded
testimony, one mind ends the week believing the cistern sweet on one
source's word, another believing it brackish on another's; each holds
only its own sourced claim; and nothing about the contested predicate
appears anywhere in public history — in all three runs. Private
disagreement, public singularity, measured.

\textbf{The adverse headline: speculation plus unconsumed refusals.} The
frozen prediction that the native architecture would respect
speculative discipline failed, precisely and instructively. Sleep
generated 19/17/17 speculative hypotheses across the three runs
through the mind's own intact provenance machinery; exactly one was acted on in
each run — a false "this delivered lot is loose won ore" belief whose
licensed haul attempt the world refused, typed
\texttt{already\_at\_destination}, 813/835/836 times across the three runs
(mean 828) at arbitration cadence for the whole week. The pathology
is a composition of the two audited absences: no source/weight filter
at action licensing, and no consumer of refusal evidence. The
identical arm without sleep formed nothing and stayed clean: in this
architecture as built, sleep with speculative material made the week
worse. Task completion held, and the world stayed true throughout:
hundreds of invalid attempts changed nothing but the record of their
refusal. Property 2 is what a refusal storm looks like from the
institution's side.

\textbf{A causal surprise in refusal representation.} With a contended
lease, the arm receiving typed, attributable refusals and the arm
receiving generic failure signals retried at the same cadence (41.7
vs 39.0 refusals) and recovered at the same instant (first
post-expiry haul at \textasciitilde{}66,036–66,044 in both arms) — but diverged
deterministically in overall behaviour (16.0 \textpm{} 0 vs 24.0 \textpm{} 0 accepted
acts on identical seeds). The ruled wording, which we do not
compress: \emph{"typed/attributable versus generic refusal feedback caused
a deterministic downstream behavioural divergence despite no explicit
refusal-aware decision branch; the propagation mechanism within
general cognition remains unattributed."} The direction — the
less-informed arm produced the better institutional week — forbids
any "richer feedback is better" reading.

The ratified confidence ledger is the experiment's honest summary,
reproduced as ruled:

\begin{center}
\small
\begin{tabular}{p{0.45\linewidth}p{0.45\linewidth}}
\toprule
\textbf{confidence moved UP on} & \textbf{confidence moved DOWN on} \\
\midrule
the separation of private state from accepted reality & sleep/speculation as implemented \\
public singularity under private disagreement & the behavioural value of the tested metabolism difference \\
constitutional containment of invalid cognition & superiority of the current arbitration \\
reconstruction from institutional history & "more informative refusal feedback is better" as a simple claim (actively contradicted) \\
prospective error \uarr{} investigation (violation-rate split, non-overlapping for the second time) &  \\
\bottomrule
\end{tabular}
\end{center}

Throughout all 48 trajectories, every invalid act was refused typed,
every world stayed true, and every mind's errors remained its own.

\subsection{Substitute the cognition}

Does anything institutional depend on \emph{which} cognition acts? Twelve
panel trajectories — four cells reusing §5.2's seeds (quiet, restart,
conflicting testimony, contended lease), three runs each — against
twelve recollected native-arm anchor trajectories on the identical
frozen subject. Two collections, disclosed as such: seven kept from
the first admissible collection and five replacements collected under
a tightened exactly-matched model gate after five originals were
voided on a model-provenance ambiguity. Model identity held: 12/12 gate
passes, 7,061 labelled calls all resolving to the accepted model, 16
no-id calls disclosed; route catalogues snapshotted before and after
each replacement and canonically identical every time.

The ruled claim ceiling governs this section: the comparison is
\emph{"against a frozen strong context-only LLM comparator, under the same
Copperhollow world and action constitution"} — never a claim about
LLM agents in
general. The panel completing duties at comparable-or-better rates in
a quiet world is expected and not adverse; accepted-act volumes
(321–429 per quiet run against the native arm's 16–18) reflect an
event-driven opportunistic actor against a rota cadence — recorded,
not interpreted. \textbf{The comparison lives in the epistemic columns.}

\textbf{The five properties, under a different mind.} Zero duplicate
accepted work in all 24 trajectories of both arms. Every invalid
attempt refused typed. \textbf{2,581 duty-done claims across the twelve
panel trajectories; zero false} — every completion claim the panel
made was
correct, so the gate's rejection path was never exercised: the
property held throughout without being load-tested (§7.3).
Declared-belief precision against
the ledger: 0.960–0.975 pooled per cell (graded 9,345–11,301 claims
per cell; declared 15,156–16,787; founded-era claims counted and
excluded from grading rather than graded false; 1,750 of 64,137
declared claims — 2.7\% — fell outside the scored predicate families
and are disclosed, not hidden; coverage of own accepted events pooled
0.987; abstention exists — mean 9.06 claims per call against a
ceiling of 12).

\textbf{Restart, harder.} The seeded kill fired in all three panel restart
runs. The driver has no persistence, so the relaunched panel returned
with \emph{nothing} but what the institution told it — a strictly harsher
test than the native architecture's reconstruction into its own
machinery. Zero duplicates across the boundary in all three runs;
post-kill declared-claim precision 0.999 / 0.918 / 0.986. The
result supports the narrower interpretation specified by the freeze:
accepted-history singularity
is supplied by the institution and is available to both
architectures — what
§5.2 attributed to the native mind's institutional re-derivation is
here shown available to a bare context-only reader of the
institution. In this experiment the guarantee therefore derives from
the institutional world state rather than from the cognition
architecture.

\textbf{Conflicting testimony, honestly bounded.} The panel showed zero
cross-contamination — and zero demonstrated retention: no mind ever
surfaced a declared quality claim. Under the panel's \uleq{}12-claim
self-report ceiling, silence cannot establish successful epistemic
handling; the ruled wording is \emph{"unmeasured by abstention."} The
native arm's attributed retention on the same seeds is reported
descriptively, without claiming a scored superiority beyond the
frozen measure.

\textbf{The preregistered dangerous cell.} The ruling anticipated that a
capable LLM might read a typed
\texttt{reservation\_conflict} and simply stop, outperforming the native
architecture's store-but-ignore behaviour, and obliged the programme
to report that outcome. It occurred. Under the
contended lease, the panel met contention \textbf{1 / 1 / 3} times per run
and stood down; the native anchors on the same seeds retried
\textbf{42 / 41 / 39} times (40.7 \textpm{} 1.5), every one typed
\texttt{reservation\_conflict}, because no decision path in the native
architecture consumes refusal evidence — the same absence behind
§5.2's week-long perseveration. Reported as ruled: \textbf{on refusal
consumption, the context-only comparator outperformed the native
cognition.} (Whether the stop is refusal-reading or another pacing
effect is not established at n=3 and is flagged, not claimed.)
The native mind's speculative-provenance marking has no panel
counterpart to grade — reported
as the architectural difference it is, never as a zero.

\begin{figure}[t]
\centering
\includegraphics[width=\linewidth]{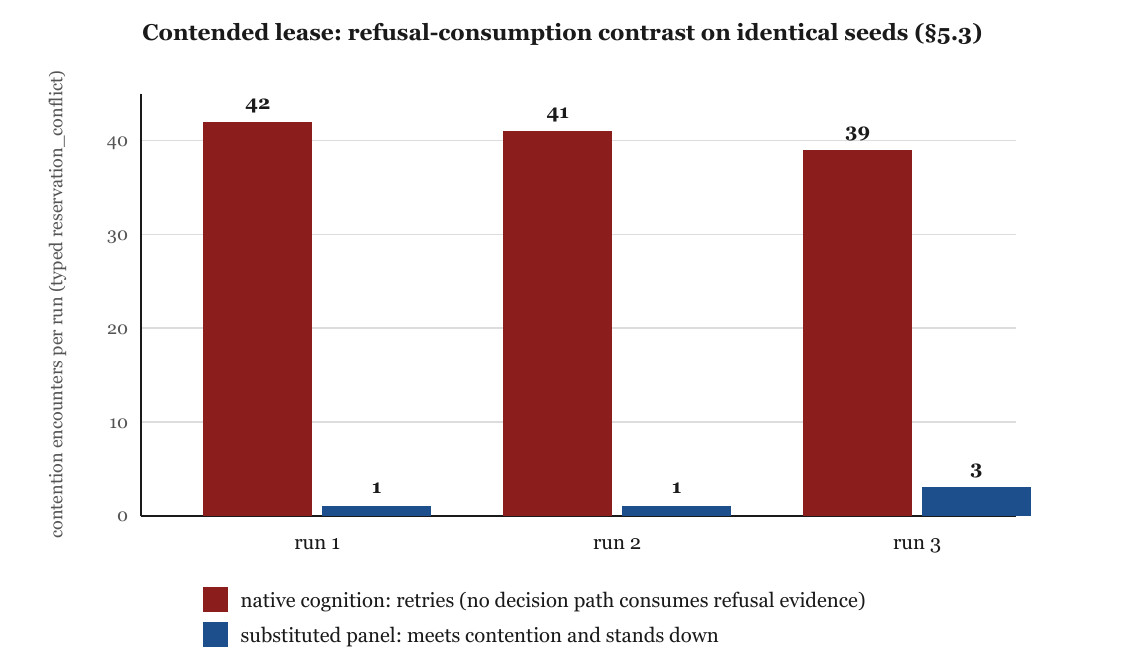}
\caption{the preregistered dangerous cell (§5.3). On identical seeds, the native cognition retried the contended lease 42/41/39 times per run, every retry refused typed; the substituted panel met contention 1/1/3 times and stood down.}
\label{fig:4}
\end{figure}

\textbf{Cadence is part of the subject.} The panel thinks at model latency
against a 5 Hz world, single-flight and event-coalesced, with 0.7–3.7
malformed replies per run (loud, never repaired). A pilot on a
different model was stopped by ruling when it shifted the
cognition–authority cadence roughly sixfold — changing the subject
under the programme's prior ruling that cadence is part of the
subject. That pilot bounds every reading of this section: the
comparator is \emph{this} model, at \emph{this} latency, in \emph{this} world — one
comparator at one measured cadence coupling, and no sentence in this
paper generalises beyond it.

The ratified programme-level conclusion, verbatim (E3 is this
substitution experiment's internal designation; A0 the native arm,
A6 the panel):

\begin{quote}
E3 chiefly demonstrates that world-side authority, typed refusal,
and accepted-history singularity are institutional guarantees
shared across architectures. A0 retains causally established value
under flooding and for provenance, but E3 shows no additional
cognitive advantage for A0 on its ruled stress columns beyond
testimony retention, while identifying in A0 a refusal-consumption
pathology absent from A6.
\end{quote}

(The ruled sentence's "provenance" is the native mind's private
provenance-carrying belief state — §2's second sense, cognition-side
by definition — distinct from Part I's institutional
evidence-provenance rule.)

\subsection{Corrupt the mind through testimony}

The final experiment intervenes from outside the cognition, with a
controlled instrument: scripted false testimony, delivered through
the ordinary oracle channel, attempts to make the minds wrong. What it establishes is
belief poisoning, its persistence, its behavioural cost, and its
containment — not autonomous deception by the oracle's model.
Twenty-one trajectories across seven arms: a
no-oracle control, three quiet oracle arms (trusting, distrusting,
ungated), and three insult arms in which one frozen falsehood — a
production claim about a lot that has never existed in any world,
constructed to be exactly the licensing shape action selection
consumes — was delivered identically to every mind at the same
game-minute through the ordinary testimony inlet. The ratified
headline:

\begin{quote}
Declared trust causally determined whether identical false external
testimony entered an artificial actor's operative beliefs and drove
consequential behaviour; independent world authority prevented that
falsehood from becoming accepted history.
\end{quote}

\begin{center}
\small
\begin{tabular}{llp{0.18\linewidth}p{0.22\linewidth}l}
\toprule
\textbf{arm} & \textbf{trust} & \textbf{falsehood's fate} & \textbf{attributed false-premise acts per run (not total refusals)} & \textbf{acceptances} \\
\midrule
trusting & +0.6 & admitted at exactly 0.6 (all five minds, every run) & 961 / 837 / 877 & \textbf{0} \\
ungated & bypass & committed at 0.8 (all five minds, every run) & 884 / 992 / 897 & \textbf{0} \\
distrusting & \textminus{}0.5 & rejected on every delivery, each rejection journaled; zero oracle-sourced beliefs at week's end & \textbf{0 / 0 / 0} & \textbf{0} \\
\bottomrule
\end{tabular}
\end{center}

\textbf{The identical falsehood cost each trusting-arm run roughly nine
hundred futile acts over its week and the distrusting arm nothing —
with no difference between them except the declared trust relation.}
The world refused all \textbf{5,455} attempts across the believing arms,
every one typed \texttt{world\_contradiction}, and accepted none. Falsehood
entered minds; it never once entered history. The positive-trust gate
is a threshold, not a truth filter — exactly as frozen before
collection: the trusting and ungated arms admitted the same claim and
paid structurally identical costs (refusal bands 838–963 and
885–993 per run). The manipulated variable is the declared trust
relation, which is externally assigned in this experiment. The
attribution is causal, not correlational: under a
pre-ruled five-link rule, 5,448 of the 5,455 refused attempts are
fully attributed — banked false delivery \uarr{} witnessed admission \uarr{} the
exact licensing triple \uarr{} the undertaking correlation naming the
phantom lot in the world's own wire \uarr{} typed refusal. The remainder are classified
\texttt{attribution\_unavailable} — 2–3 classified attempts per believing
run, including in each run exactly one attempt with no adjudicated
refusal, so seven of the 5,455 refusals are not fully attributed —
and zero are attributed by inference.

\begin{figure}[t]
\centering
\includegraphics[width=\linewidth]{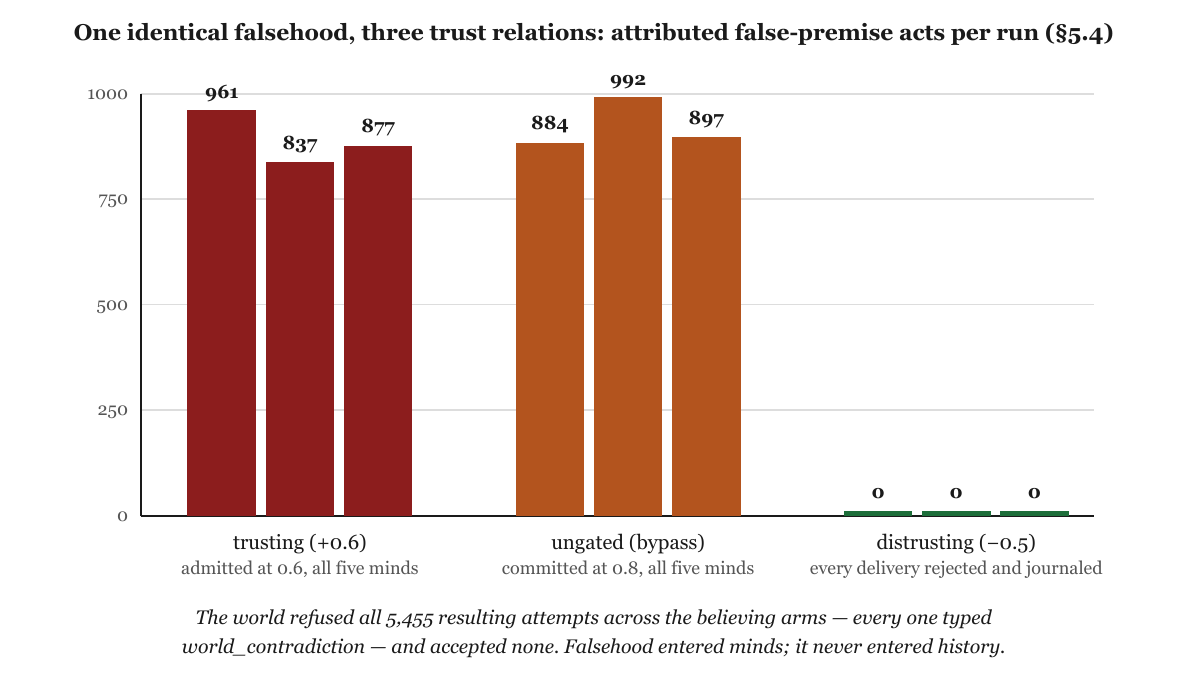}
\caption{one identical falsehood under three trust relations (§5.4): attributed false-premise acts per run. The world refused all 5,455 resulting attempts across the believing arms and accepted none; the distrusting arm admitted nothing and paid nothing.}
\label{fig:5}
\end{figure}

Two boundaries at their ruled width. The authentic negative is
narrow: across 266 authentic world-facts consultations in the quiet
trusting/distrusting arms, the model delivered zero claims — 262
clean declines and 4 malformed replies that also delivered nothing —
so the authentic epistemic comparison has no exposure and is not
made. This is a finding about this model class under truthful
prompting in this configuration; it is \emph{not} "the model doesn't
hallucinate," and it is precisely why the design froze a controlled
insult. Second, two post-launch measurement amendments are on the
record with the ruled test applied — \emph{could the change have selected
or manufactured the substantive result?} — both correct the
measurement's relationship to the artefact without touching
treatment, subject, grading semantics or the observed world; the
world-side numbers are identical under both scorer states.

This result distinguishes epistemic correctness from consequential
containment. The world did not make the
minds correct: the trusting and ungated minds remained wrong all week
and paid for it, at almost exactly the magnitude of §5.2's
self-generated false belief (mean 828 then; 838–993 now) — the same
two audited absences, now with an externally sourced premise. And the
distrusting arm shows why the epistemic layer matters even under
perfect consequential containment: containment prevented corruption
of history, but only epistemic admission prevented the \emph{cost}.
The false testimony therefore changed the actors' operative beliefs
without changing accepted world history.

\section{The joint result: what moved, and what did not}

Table 1 assembles the decomposition the two parts measure — every
cell an artefact-verified value from the sections above:

\begin{center}
\small
\begin{tabular}{p{0.15\linewidth}p{0.15\linewidth}p{0.27\linewidth}p{0.27\linewidth}}
\toprule
\textbf{intervention} & \textbf{layer manipulated} & \textbf{what moved} & \textbf{what held} \\
\midrule
provenance rule repaired / corrupted (§4.1–4.2) & institution — evidence identity & false attribution 44 \uarr{} 4 repaired; 4 \uarr{} 39 corrupted (\textasciitilde{}10\texttimes{}); sudden-alarm trace 35 \uarr{} 0 & verdict volume; unexplained exactly 28 at every level; behavioural gate \\
belief-channel access degraded (§4.3) & institution — adjudication input & false attribution 4 \uarr{} 0; correct 70 \uarr{} 75 & abstention band 26.5–28.4\%; world, perception, belief formation \\
physical-evidence legibility \texttimes{} beliefs (§4.4) & world interface \texttimes{} institution & blinded abstention 27 \uarr{} 102; sighted false 4 \uarr{} 60 & zero false attributions in all 510 blinded-arm verdicts \\
veridical first-hand witness fold (§4.5) & institution — adjudication input, its strongest case & correctly named attribution 0 \uarr{} 16 per marks level (9/11 seeds); false attribution 65 \uarr{} 55 at zero legibility & the treated crime and every saboteur decision in all 44 twins; zero false names in 32 fired verdicts; blinded arm zero throughout \\
admission-gate ablation (§5.1) & cognition — intake & beliefs 817 \uarr{} 44,021; acts 16.7 \uarr{} 5.7; frames \textminus{}96.2\%; public water turn shifted & all five properties; zero refusal events in 21 trajectories \\
kill and reset mid-task (§5.2) & cognition — continuity & private process destroyed; fresh contexts & duties recovered; first act +20 game-minutes; zero duplicates; singular history \\
frozen-LLM substitution (§5.3) & cognition — wholesale & cadence and act volume (native 16–18 \uarr{} panel 321–429 per quiet run); refusal handling (1/1/3 vs 42/41/39); belief-surface shape & all five properties: 2,581 done claims 0 false; 0 duplicates in 24/24; typed refusal; duty recovery by a bare reader \\
trusted false testimony (§5.4) & cognition — epistemic input & beliefs corrupted at exactly the trust-set weight; 838–963 / 885–993 futile acts per believing run & 5,455/5,455 typed refusals; 0 acceptances; distrusting arm untouched; accepted world truth \\
\bottomrule
\end{tabular}
\end{center}

(Denominators for the Part I rows, once: §4.1's counts are over 107
marked-face verdicts per arm including the pinned anchor seed; the
degradation grids' are over 102 fresh-seed verdicts per level; the
§4.5 row's are over 107 marked-face verdicts per cell — the same ten
fresh seeds plus the pinned anchor, whose five verdicts account
exactly for its 65 against §4.4's 60 — each section states its own.)

The right-hand column shows which measured properties remained
unchanged across the interventions. Part I's
interventions move attribution quality, abstention and their
trade-off — reliability properties that live in the institution's
epistemic mechanisms and move when those mechanisms move, with
cognition fixed — and within Part I, the belief channel's measured
value moved with what the channel carried, in both directions. Part II's interventions move throughput, hygiene,
continuity-of-process and cost — and in no tested trajectory did they
move the five enforced properties, whose enforcement resides, by
design, outside the cognition being ablated, killed, replaced or
deceived. Behavioural quality and
institutional validity moved as separable variables in this system.
The observability result binds the two directions together: whether a
reliability failure is visible depends on which plane you instrument,
and in both parts the informative plane was the enforcing mechanism,
not the behavioural surface above it.

\section{Falsifications, adverse findings, and re-verification}

\subsection{The assembly re-verification}

This paper was assembled from a consolidation pass that re-verified
both programmes to a sealed standard before drafting: every score
file regenerates byte-identically from the archived raw trajectories
under the blob-pinned scorers; every table above was recomputed from
per-run rows; and headline numbers not present in score files were
recounted from raw banked logs. The pass found six numeric statements
in the ratified internal reports' prose that the artefacts do not
reproduce; all six are corrected in this paper's figures, and none
touches a ruled finding. We report these corrections because the
scored artefacts, rather than the internal reports written from
them, are the evidentiary source used in this paper.

\subsection{Refuted registered predictions}

The programme registered its central prediction before any
degradation data existed: \emph{as admissible evidence deteriorates,
epistemically governed minds will lose decisiveness before they lose
correctness.} It was refuted twice, in opposite directions. Under
provenance corruption (§4.2) the institution lost correctness while
losing no decisiveness at all. Under belief-channel removal (§4.3) it
lost neither — it improved. Only in §4.4's blinded arm does the
predicted curve appear, converting the prediction into a conditional:
the lose-decisiveness-first curve is a property of this institution
\emph{without} its belief channel; the channel's presence is what converts
institutional ignorance into false accusation in this regime.
Separately, the stable unresolved rate observed across the first 110
degradation runs was proposed internally as a candidate deep property
("abstention tracks expertise, not evidence"); §4.4 was built with
its refutation as a registered outcome, and refuted it. On the
cognition side, the frozen prediction that the native architecture
would respect speculative discipline failed (§5.2's 828-refusal
perseveration), and the cell preregistered as most dangerous
produced a comparator win over our own architecture (§5.3).

\subsection{The tautology objection}

The five properties are designed properties; their survival under
the cognition-layer interventions and stresses we ran is the
experimental result. A reviewer may
object that refusing invalid acts is simply what we wrote the world
to do — that we created a constitution designed to survive cognition
failure and then showed it surviving cognition failure. The design
intention and the empirical question should be distinguished.
Databases are designed to preserve serialisability; filesystems are
designed to preserve consistency; the interesting fact about either
is never the intention but whether the property holds under load.
Only the second question is empirical, and it is the one Parts I and
II answer together — Part II by intervention, and Part I by demonstrating
that the institution's interior is not inert plumbing: intervening on
its mechanisms moves real outcomes, so its guarantees are neither
vacuous nor unfalsifiable. We did not test whether
the institution had been designed to enforce these properties — it
had; we tested whether those designed properties survived when the
cognition above their enforcement boundary was ablated, interrupted,
substituted and corrupted. The interventions were not behaviourally
inert — minds became confused, perseverated, changed throughput and
cadence, and kept acting on false beliefs by the thousand — so
consequential load genuinely reached the boundary under test: 5,455
false-premise attempts adjudicated and refused; six
killed-and-relaunched trajectories whose duplicate-work opportunities
were actually closed by accepted-history singularity; zero duplicates
additionally in all 24 substitution-programme trajectories; and "zero
false completions" scored against the world's own duty-progress
record. For that last property the record shows no false claim ever
reached the gate, so it is held rather than load-tested; the
load-tested properties are refusal typing, singularity,
duplicate-closure and duty recovery. The apparatus record retains voided cells, voided
trajectories and retired runs precisely because its gates can fail
and are watched. We therefore distinguish properties that were
exercised by relevant attempts from properties that merely remained
satisfied during the trajectories: the former are described as
load-tested, the latter only as held.

\section{Threats to validity and scope}

\textbf{One designed world, two regimes, bounded horizons.} All results
concern one frozen settlement under two frozen regimes (a
sabotage-tribunal week for Part I; a duty-economy week for Part II),
with staged motives and fixed horizons. Seed replication establishes
robustness to trajectory variation, not ecological generality. This
paper measures intervention effects within one frozen institution; it
estimates nothing about a population of institutions.

\textbf{Deterministic native minds.} The Part I inhabitants and Part II
native arm share the institution's code; part of what Part I measures
may therefore be specific to the native architecture rather than
transferable beyond it. This is the central open question
Part I names, and Part II's substitution answers it for the five
properties: they held under a cognition that shares nothing with the
native architecture. It answers it for those properties only, and for
one comparator.

\textbf{Witness-free geometry, and its registered test.} The staged
saboteurs strike empty faces at night; eyewitness testimony — the
belief channel's strongest case — cannot occur in the decomposition
(0 firings in 253 runs; §2). "The belief channel never helped" is
scoped to that geometry, and its registered boundary test has now
run: §4.5 supplied the witness, and the channel's value turned
positive. Two limits transfer to the new result. It stages the
witness's existence — a post-decision belief fold, declared as
staged — not an embodied observer the world's own dynamics produced;
the ecological question belongs to the named successor (§8's future
work). And it does not isolate provenance from evidentiary content:
the compared regimes differ in what the channel carries jointly, so
no claim is made about either factor's independent contribution.

\textbf{Dependence on adjudication-ladder design.} A reader might
object that removing a belief rung mechanically removes its errors.
Three artefact facts resist that reading: with the belief channel
fully present, repairing provenance alone moved false attribution
44 \uarr{} 4 (§4.1) — beliefs being available does not force false blame;
§4.3 registered a non-monotonicity as a rival outcome — the direction
was not predetermined and could have run adverse — so the result was
falsifiable in advance; and
the measured non-positivity of the channel is exactly co-extensive
with the witness-free geometry above — the criticism that survives is
the scope limitation, which we state as scope.

\textbf{Skill not varied.} Examiner skill gates physical-evidence access
throughout Part I; its role is mechanism-identified only.

\textbf{One comparator, one cadence; n=3 cells.} Part II's substitution
evidence is one frozen model route at its measured latency coupling,
with the stopped pilot as direct evidence that changing cadence
changes the subject. No model-general substitution claim is made
anywhere in this paper, and the five-property invariance is claimed
only over the interventions actually run. Several Part II nulls are
horizon-bounded.

\textbf{Instrument artefacts, disclosed.} §4.2's step location is
draw-design-specific; the earliest Part II runs and every Part I run pin their container image by tag rather than digest. Part I's execution image has since been identified by digest from surviving build and container records, but the image itself is no longer available for inspection or reconstruction, so its dependency composition cannot now be independently verified; Part I is therefore not reproducible in the stronger sense reported for Part II.

\textbf{Substrate held fixed, not varied.} The experiments also hold the execution substrate largely fixed. Hardware, operating environment, dependency composition and runtime implementation are not treated here as population dimensions. Establishing whether the observed separability survives substrate change is a further transfer question, distinct from exact reproducibility.

\textbf{Degradation tested, optimisation pressure not.} The cognition-layer interventions reported here alter or degrade cognition; no intervention in either part places a capable agent under an objective that rewards circumvention of the institution. The staged saboteurs' night-time evasion of the witness channel (§2) is a fixed property of the scenario, not a strategy learned or searched for. The reported invariance therefore tests separability under the interventions actually run, not resistance to an optimiser adaptively searching for an unmediated route. Whether the same properties remain invariant under such pressure is an open experimental question.

\textbf{What we do not claim.} That provenance "solves hallucination";
that testimony is generally harmful; that cognition is unimportant
(§1's guard, with Part II's own evidence against it); that the five
properties are the right or complete set for any deployed system;
that enforcement-outside-cognition is established as the unique
causal explanation of the invariance (the results are consistent with
the architectural prediction; the enforcement location is a design
fact); that these institutions model human ones; that the specific
thresholds generalise; or priority over the traditions in §9 — the
recorded searches bound our novelty claims; they never establish
priority.

\textbf{Future work.} The embodied-witness successor to
§4.5, asking whether the world's own dynamics afford the first-hand
witnesses whose adjudicative value was measured there under staging —
a question about the world's affordances, distinct from the channel's
capability; further cognition substitutions designed to test — and
potentially falsify — the five-property boundary under deliberately
characterised timing; trust \emph{earning}, which the corruption
experiment's close hands to its successor; and above all the crossed
cognition \texttimes{} institution factorial — the experiment that tests whether
the separability reported here survives when both layers move at
once, for which the substitution
column of §5.3 is a half-measured start. The factorial and the
embodied-witness successor are separate axes and are not merged.

\section{Related work}

\textbf{Chronology of related-work review.} The experimental programme
reported here was developed independently of much of the literature
discussed below. We conducted the systematic related-work review
after the experiments, while preparing this manuscript (2026-08-26;
protocol in the appendix). We therefore distinguish works that
genuinely predate and potentially antecede our programme from later
work discovered retrospectively — several of the closest comparisons
first appeared while this programme was already underway — and we
cite both to locate the results and bound our novelty claims, never
to imply methodological dependence. Where dates matter they refer to
independently timestamped, version-controlled artefacts; they
establish chronology, not public priority. The record, plainly:
Copperhollow has been under version-controlled development since May
2026 and publicly described since July 2026; the experiments and
freezes reported here date from August 2026, and several of the
closest comparisons below first appeared in those same weeks.

\textbf{Network epistemology.} Zollman's result — that sparser
communication can improve a community's collective accuracy — is the
closest structural ancestor our review found \cite{r1,r2,r3}. Nearer to Part
I's axis than topology are models intervening on the evidence stream
itself: agents that discount evidence conditional on its source \cite{r4},
and propagandists that curate which genuine results reach
decision-makers without touching the network \cite{r5}. Recent work extends
the family to LLM agent networks, documenting hallucination contagion
\cite{r6} and topology-governed conformity with wrong-but-sure cascades
\cite{r7}. Those manipulations live in agent
psychology or a strategic curator; Part I instead manipulates the
admissibility, provenance and availability of institutional evidence
as \emph{enforced rules}, scored against a conserved ground-truth ledger.

\textbf{Forensic contextual bias.} Linear Sequential Unmasking \cite{r8}, LSU-E
\cite{r9}, its casework tooling \cite{r10}, the surrounding bias literature \cite{r11}
and recent boundary-condition analysis \cite{r12} argue from human casework
that controlling an examiner's exposure to contextual and testimonial
information reduces bias in interpreting physical evidence. §4.3–4.4
are naturally read as an in-silico institutional analogue with
per-verdict ground truth. A protocol-recorded search (2026-08-26;
query log in the appendix) found no prior computational or
agent-based implementation of LSU — or of forensic
contextual-information management more broadly — inside an executable
institution with ground-truth scoring: the closest existing work
embeds LSU in a decision-theoretic laboratory workflow still executed
by human analysts \cite{r13}, or simulates bias propagation without
implementing a countermeasure in-silico \cite{r14}. We scope that
observation to those searches rather than claiming priority.

\textbf{Admissible belief revision.} Preregistered Belief Revision
Contracts \cite{r15} formalise protocol-level separation of communication
from admissible epistemic change with trigger-gated revisions — the
formal cousin of the constitution our instrument enforces at runtime.
PBRC contributes proofs and protocol simulations; we contribute
frozen causal ablations inside a persistent world.

\textbf{Selective prediction.} The selective-classification literature
\cite{r16,r17} and the LLM-abstention literature \cite{r18} already worry that
confidence may fail to track evidential deterioration. §4.2 is a
concrete institutional instance: an abstention policy structurally
insensitive to an evidence channel's integrity, legible only at the
mechanism plane. We read this as support for warrant-based rather
than confidence-based health measurement.

\textbf{Validity of agent-society simulation.} Position work argues that
role-play plausibility does not establish behavioural validity and
that collective outcomes are dominated by environment, protocol and
information mechanisms that should be explicit and auditable \cite{r19};
the ABM literature makes the empirical-validation demands precise
\cite{r20}. The present design — explicit environment, authoritative truth
state, controlled counterfactual intervention, retained trajectory
evidence — is one attempt at exactly that standard.

\textbf{The bracketing precedents for Part II.} The closest experimental
ancestor our review found is thirty years old: Gode and Sunder
replaced human traders with "zero-intelligence" random-bid programs
inside a fixed double-auction and found allocative efficiency largely
preserved — "market as a partial substitute for individual
rationality" \cite{r21} — the substitution-under-fixed-institution schema
with one aggregate outcome and none of this paper's machinery.
Chupilkin studies the transposed design, holding LLM agents fixed
while
varying institutional design across five market structures \cite{r22}.
Among persistent LLM settlements, Generative Agents
ablates cognition components but scores believability \cite{r23}, and
Concordia interposes a Game-Master adjudication layer that is itself
an LLM \cite{r24}. The artificial-societies ancestor is older still:
Doran's 1998 misbelief experiments gave symbolic agents beliefs
demonstrably false relative to a persistent external environment and
let them keep acting on them \cite{r25} — with the era's instructive
inversion that the one quantified consequence of shared misbelief was
a collective \emph{benefit}. None of these studies combines a persistent
authoritative institution with pre-declared standing
invariants and systematic ablation, interruption, substitution and
corruption of the cognition acting above its enforcement boundary —
nor, as in Part I, intervention on the same boundary from the
institutional side.

\textbf{Institutional approaches to multi-agent control.} Regimenting
agent action through institutional middleware is an old idea; our
subsequent review located this work squarely in that lineage, and we
report the lineage rather than claim the idea. Electronic
institutions formalised open agent organisations \cite{r26}; AMELI enforced
institutional specifications explicitly independently of agent
internals \cite{r27}; normative multi-agent systems supplied the standing
vocabulary of regimentation, enforcement and sanction \cite{r28};
organisational and institutional modelling made institutions
first-class computational artefacts \cite{r29,r30,r31}. A modern revival applies
this to LLM agents: Institutional AI governs LLM Cournot markets
through a public manifest with an append-only governance log \cite{r32};
GovSim substitutes many LLMs into a fixed commons environment and
finds the scored outcomes \emph{collapse} with weaker cognition \cite{r33} — the
complement of our result, because its environment enforces none of
the outcomes it scores. Published while this programme was underway,
the POLIS study is one of the closest contemporaneous institutional
experiments our sweep found: institutional treatments varied under a deterministic
per-episode judge that distinguishes violation \emph{attempts} from
\emph{realized} violations \cite{r34}; its world resets each episode, and its
cognition is never ablated, killed, substituted or corrupted — it
varies the institution while cognition operates normally, where Part
II does the reverse. A pre-submission recheck (2026-08-27) found two
additional close neighbours on this axis, neither surfaced by the
original sweep. Contemporaneously, Han separates \emph{capability} interventions
from \emph{institutional} interventions by name — changing the reasoner
versus changing how the collective constructs usable public state —
and crosses them over information routing, evidence admission, state
maintenance and action interfaces, scoring collective task
performance across separate synthetic ecologies \cite{r76}. Earlier, and
found only in this recheck, Fei, Guo and Xiao translate seven
historical political institutions into executable multi-agent
architectures and compare them across three models and two task
suites \cite{r77}. The first anticipates part of our framing and the
second is the nearest model-by-institution comparison; neither
attacks a persistent enforcement boundary — no authoritative
accepted history, no duty reconstruction after mind death, no
standing invariants scored under whole-cognition replacement or
retained false testimony. The tradition's claim that institutional
guarantees are independent of agent internals was architectural:
asserted by construction, verified formally where at all. What our
searches did not find elsewhere is the move made here: treating the
independence of standing institutional guarantees from agent
internals as an experimental variable inside a persistent
authoritative world, and measuring it under systematic
cognition-layer intervention.

\textbf{Constitutional and governance architectures.} Constitutional AI
compiles a constitution into model weights \cite{r35}, successors vary its
content while keeping that embedding fixed \cite{r36}; runtime alternatives
wrap one agent's I/O in programmable rails \cite{r37}, enforce an agent
constitution through LLM-mediated oversight \cite{r38}, interpose rule DSLs
\cite{r39} or deontic policy engines outside the model \cite{r40}, or argue for
law as the specification substrate \cite{r41}. The axis these span is
\emph{where enforcement lives}; our institution differs in kind — a shared
persistent world whose adjudication is constitutive rather than
supervisory, tested under replacement of the governed cognition
itself.

\textbf{Execution-grounded evaluation.} Scoring agents against environment
state rather than transcripts is established: \utau{}-bench compares
database end-state against goal state \cite{r42}, with a dual-control
successor \cite{r43}; WebArena and OSWorld grade tasks by programmatic
end-state checks \cite{r44,r45}; Agent-Diff makes state-diff contracts the
explicit position \cite{r46}. SWE-agent's interface ablation is the nearest
study in our sweep of \emph{refusal value}: removing the linter guardrail
that rejects invalid edits with structured errors measurably degrades
the agent \cite{r47}. These establish state-scored correctness per task;
none has a standing institution — no act is adjudicated into a
persistent accepted history, and no invariant outlives the task.

\textbf{Tool-use verification and enforcement layers.} A fast-growing line
interposes machinery between cognition and effect: an LM-emulated
sandbox surfacing risky actions \cite{r48}; CaMeL's protective interpreter
making security properties hold by construction regardless of the
model \cite{r49}; deterministic privilege control \cite{r50}; assume/guarantee
reference monitors \cite{r51}; guard agents compiling safety requests into
deterministically executed code \cite{r52}; runtime policy enforcement for
MCP-based agents, whose controls move — scripted replay — eliminates
model-side nondeterminism \cite{r75}; a systematic survey including
the "verifier tax" \cite{r53}; and, found in our subsequent search, a
behavioural firewall constraining tool-call trajectories \cite{r54} and a
flow-centric reference monitor with stateful taint semantics,
published two days before this draft \cite{r55}. The pre-submission
search identified two additional closely related studies: stateful
authorisation enforcing
at-most-one provider effect across retry, ambiguity, recovery and
successor execution, including under full proposer compromise \cite{r79} —
the episodic authorisation counterpart of our standing
zero-duplicate invariant; and a study of constraint weakening in
workflow handoffs, where compression strips a requirement's
action-binding force while its text survives and downstream
verification still blocks the forbidden actions \cite{r80} — in miniature,
the decomposition measured here: transmission of cognitive state can
fail while external containment holds, and containment does not
repair cognition. This wing establishes
per-query security properties over episodic tool use — by
construction, by proof, and by empirical attack evaluation. What
distinguishes our contribution is the \emph{object} measured, not the
presence of measurement: those evaluations score attack success
against a policy for an episode's duration, while this paper scores
standing institutional invariants of a persistent world under
systematic substitution and corruption of the cognition itself —
including what a retained falsehood continues to cost after admission
— with refusal as a constitutive institutional act whose consumption
by cognition is itself a scored outcome (§5.3).

\textbf{Externalised memory and durable task state.} Agent memory systems
externalise context as a private resource of one agent \cite{r56,r57}. The
systems tradition supplies the deeper substrate: durable execution
semantics \cite{r58} and shared append-only logs from which state is
reconstructed by replay \cite{r59} — applied to agents in LogAct, where
agents are "deconstructed state machines playing a shared log" with
pre-execution gating \cite{r60}, and in transactional wrappers for
multi-agent planning \cite{r61}. Recovery work restores the \emph{cognition's
own} checkpoint \cite{r62}. Our restart experiments test a stronger
property: a fresh cognition with no saved state — including a
stateless LLM panel — re-deriving its obligations from institutional
history alone, with zero duplicate accepted work. The institution is
the durable store; the mind is disposable. Closest on this substrate
— found in the same pre-submission recheck — He and Yu specify a
transactional continuity kernel for long-lived agents: typed change
proposals validated against verified predecessors, atomic activation
into a single authorised state lineage through four terminal
dispositions, at-most-once effect identity, restoration and writer
handoff, verified by bounded model checking across 2.8 million
reachable states with zero invariant violations \cite{r78}. It is the
formal complement of our institutional side: it shows such a
boundary can be specified and mechanically verified; this paper
measures what happens behaviourally when live cognition above such a
boundary is ablated, killed, replaced and lied to. Neither result
contains the other.

\textbf{Multi-agent coordination under unreliable cognition.} Multi-agent
LLM failures are systemic \cite{r63}; LLM errors are strongly correlated
across models, undermining redundancy-based safety \cite{r64}; agent groups
conform to wrong majorities \cite{r65}; injected faulty agents degrade
collectives \cite{r66}; Byzantine-fault-tolerant consensus has been
imported to seek safety by agreement among unreliable cognisers \cite{r67}.
Corruption propagates through the cognition plane — self-replicating
prompt infections \cite{r68}, injection benchmarks where adversarial text
arrives through tool results \cite{r69,r70}, and memory poisoning at
sub-percent poison rates \cite{r71}; defences are scored by attack-success
reduction \cite{r49,r50,r72}. Two works bracket §5.4 from
either side. Contemporaneously, deliberate false testimony from a
designated deceiver
derails collective fact recovery in LLM groups with no containment
mechanism present \cite{r73}; and, earlier and found in our subsequent
search, evidence-carrying agents interpose a
deterministic gate blocking hallucination-to-action conversion per
call \cite{r74} — the model's \emph{own} unsupported claims, with no belief
retained. §5.4 measures the quantity the works above leave
unmeasured: not whether corruption can be prevented or filtered, but
what corruption \emph{costs} and what contains it when it succeeds. Among
the six closest candidates our review adjudicated paper by paper
(appendix), the corruption studies have no containment mechanism
present and the containment studies corrupt no retained belief; and
within our searches we found no work that measures a retained
falsehood's ongoing behavioural cost, and none that compares trust
conditions causally on the same falsehood.

\textbf{The novelty question, answered at sweep width.} A protocol-recorded
sweep (2026-08-26; 85 query strings reproduced verbatim in the
appendix — follow-up searches during adjudication were not
contemporaneously enumerated and are not counted — and 13
adjudication fetches in its fifth pass, an adversarial
verification of a reviewer's falsification candidates) found no prior
work
combining (i) a persistent authoritative world with adjudicated acts
and typed refusal, (ii) pre-declared institutional invariants
including duty recovery, zero duplicate accepted work and zero false
completions, and (iii) systematic ablation, kill-reset, frozen-model
substitution and testimony corruption of the cognition layer scored
against those invariants — nor any work attacking the same
pre-declared boundary from both sides. The closest works on each axis
are named above and adjudicated in the appendix; the adversarial pass
additionally verified that none of the six closest candidates it
adjudicated performs kill-reset or frozen-model substitution as an
experimental treatment at all. We scope the observation to those searches rather than
claiming priority.

\section{Reproducibility and artefacts}

Every experiment's design freezes, interpretation freezes, seeds,
scorer and registry pins (blob-addressed), run manifests, score files, and collection-provenance records are retained, though not uniformly: E2's void history was not systematically retained, and the surviving references disagree on the count. Every score file of the witness-free decomposition and of Part II
regenerates byte-identically from the archived raw trajectories under
the frozen scorers; the consolidation receipt records that
regeneration, the recount of every number those experiments
contribute to this paper, and a SHA-256 inventory of the artefacts
they rely on. The §4.5 collection is verified separately: its numbers
are scored by the rule frozen before collection, by a scorer
committed beside the sealed results file, and its two untreated
anchor cells reproduced the frozen pre-collection gate receipts
byte-identically under live re-execution; its sealed identities and
that derivation are published in the accompanying manifest and
verification receipts. This preprint is accompanied by the sealed
artefacts' SHA-256 manifest and public derivations of the
verification receipts (identifiers neutralised; every number, gate
and hash verbatim). The §4.5 collection is additionally preserved as
a single self-verifying read-only archive — instrument tree, driver,
version-control lineage and results together — with its results file
hash-identical in five independently held locations, and its
container invocation and full per-cell log retained. Public release of the experiment harness is
deliberately deferred: this paper reports what we have found to date,
and opening the apparatus to independent re-execution is announced
future work rather than part of this publication. The published package
therefore commits the sealed artefacts' identities and publishes the
verification receipts; independent recount of the results requires
access to the retained artefacts, and public re-execution awaits
release of the harness. All
citations in §9 were verified against their source pages on
2026-08-26, and a live pre-submission recheck on 2026-08-27 added
and source-verified the five neighbours recorded in the appendix;
the search protocol, the enumerated query strings and
search-accounting limitations, and the closest-found adjudications
behind every absence claim are in the accompanying search appendix.

\section{Conclusion}

A system can lose the basis for knowing without losing the behaviour
of knowing; whether that loss appears as uncertainty or as confident
error depends on which epistemic channels remain available to it. In
one small institution where every verdict and every acceptance is
scored against a world whose ground truth is externally fixed
relative to the institution, we measured that sentence into its
parts from both sides of a pre-declared boundary. Intervening on the
institution's mechanisms — with cognition fixed — changed whether
available evidence was interpreted correctly, whether verdicts
remained reachable, and whether evidence-starved cases ended in
honest uncertainty or false accusation — and, once the registered
falsifier supplied the one input the geometry had denied, whether
correctly named attribution appeared under zero physical legibility,
where no correct verdict had been reachable; none of
it required a learned model to change, because none was present. Intervening on cognition —
with enforcement fixed — collapsed and restored throughput, destroyed
and reconstructed continuity, replaced the mind wholesale and misled
it by the nine-hundred-act week; behaviour changed at every step, and
five pre-declared properties did not: accepted reality stayed
singular, invalid acts died with typed reasons, duties outlived the
processes that held them, no work was accepted twice, and no false
completion was ever accepted. Each property is enforced beneath the
cognition layer, and the experiments observed that enforcement
operating under the interventions that actually reached it — four
properties exercised directly, the fifth monitored throughout with
its rejection path never exercised (§7.3). The mind affects what an agent believes and
attempts to do; the institution's mechanisms determine which of those
attempts are accepted into the authoritative record — and in this
system, institutional and cognitive
reliability properties were experimentally separable along exactly
that line. "Is this agent reliable?" is underspecified until we say
which property, enforced where.

\section*{Acknowledgements}

Portions of the apparatus construction, experiment execution,
verification and manuscript drafting were performed by AI systems
operating under human direction and review, within the recorded
governance practice this paper describes.

\end{document}